\documentclass[a4paper,12pt]{article} 
\usepackage{amsmath,amssymb,graphicx,amsthm,psfrag} 
\numberwithin{equation}{section} 
\newtheorem{thm}{Theorem}[section]

\newtheorem{prop}[thm]{Proposition}

\newtheorem{lem}[thm]{Lemma}
\newtheorem{rem}[thm]{Remark}

\newcommand{\Z}{\mathbb{Z}}

\newcommand{\st}{\textrm{ s.t. }}
\newcommand{\Pb}{\mathbb{P}}

\newcommand{\E}{\mathbb{E}}

\newcommand{\e}{\varepsilon}

\newcommand{\Card}[1]{|#1|}
\newcommand{\Cset}{\mathbf{C}}
\newcommand{\C}{{\cal C}}

\begin{document} 
\title{\em Information Loss in Coarse Graining of Polymer Configurations via
Contact Matrices}
\author{Patrik L. Ferrari\thanks{Zentrum Mathematik, Technische Universit\"at
M\"unchen, D-85747 Garching, Germany\newline \vspace{6pt} e-mail: ferrari@ma.tum.de} \, and Joel L. Lebowitz\thanks{
Departments of Mathematics and Physics, Rutgers University,
Piscataway, New Jersey\newline e-mail: lebowitz@math.rutgers.edu}}
\date{December 3, 2002} 
\maketitle

\begin{abstract}
Contact matrices provide a coarse grained description of the
configuration $\omega$ of a linear chain (polymer or random walk) on
${\Z}^n$: $\C_{ij}(\omega)=1$ when the distance between the position
of the $i$-th and $j$-th step are less than or equal to some distance
$a$ and $\C_{ij}(\omega)=0$ otherwise.  We consider models in which
polymers of length $N$ have weights corresponding to simple and
self-avoiding random walks, SRW and SAW, with $a$ the minimal
permissible distance. We prove that to leading order in $N$, the
number of matrices equals the number of walks for SRW, but not for SAW.
The coarse grained Shannon entropies for SRW agree with the fine
grained ones for $n \leq 2$, but differs for $n \geq 3$.
\end{abstract}

\section{Introduction}\label{sect0}
The use of coarse grained descriptions is essential for systems with
many degrees of freedom. The choice of the coarse grained variables is
dictated by the nature of the system and by the questions of interest.
One is then interested in the amount of information lost in the
coarse graining, at least in some statistical sense~\cite{Bill}.

\begin{figure}[t]
\begin{center}
\psfrag{omega1}[0][0][1.6]{$\textrm{Random walk }\omega_1$}
\psfrag{omega2}[0][0][1.6]{$\textrm{Random walk }\omega_2$}
\includegraphics[angle=0,height=5cm]{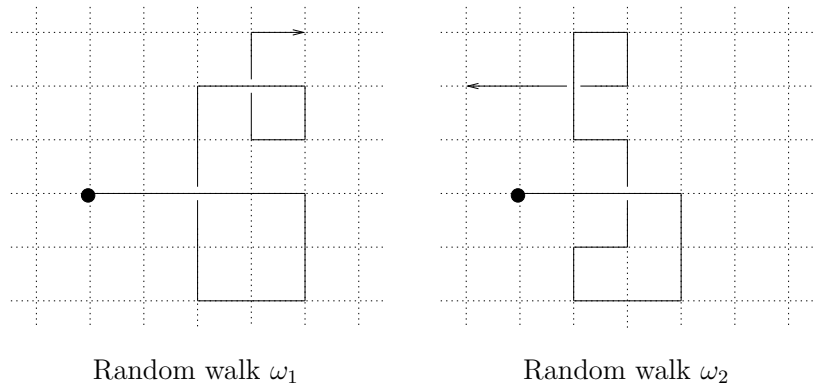} 
\caption{The two random walks $\omega_1$ and $\omega_2$ have the same contact matrix \newline because they have the same self-intersections: $2-10$ and $13-17$. \label{figex}}
\end{center}
\end{figure}

In this paper we shall study this question for simple models of  polymers, large molecules consisting of a linear sequence of $N$ monomer units.  A reduced description of this system can be based \cite{HCD, LS, Leb, VKD} on associating to each polymer configuration a connectivity or contact matrix $\C$, such that $\C_{ij} = 1$ or $0$ depending on whether the distance between the position of the $i$-th and $j$-th monomer, $\omega(i)$ and $\omega(j)$, is smaller or bigger than a certain specified value $a$,
\begin{equation}\label{eq1.1}
\C_{ij}(\omega)=\left\{\begin{array}{l} 1 \quad \textrm{ if
} |\omega(i)-\omega(j)| \leq a, \quad i\neq j,\\ 0 \quad \textrm{ otherwise.}
\end{array}\right.
 \end{equation}
This coarse grained (see Figure~\ref{figex}) 
representation of the structure of proteins is often used for
numerical studies of protein folding.  The very minimalist nature of
this representation permits a rapid first search for a protein's
native structure in terms of its contact matrix.  In fact, knowledge of
the contact matrix can predict many features of the vibrational
spectra of certain proteins \cite{Bahar2,Bahar}.  This makes it
important to have information about the relation between the space of
contact matrices and that of the proteins they represent.  

To answer the question of how much information about a polymer is
retained by its contact matrix, we consider an idealized version of
the geometrical structure of a polymer in which the monomers occupy
sites on the $n$-dimensional lattice $\Z^n$, and consecutive monomers are on nearest neighbor lattice sites.

We compare the Shannon entropy after coarse graining, $S_C(N)$, with
the Shannon entropy without coarse graining $S(N)$. To quantify the
loss of information in the coarse graining we consider
$\delta_N=S_C/S$. We prove, for SRW on $\Z^2$, that $\delta_N$ goes to
one as $N$ becomes large, showing that the relative loss of
information $(S-S_C)/S$ vanishes.
This is a consequence of the recurrence of SRW on $\Z^2$. Moreover we provide
some bounds on finite size corrections (see (\ref{eq2.2b})). On the
other hand, for SRW in $\Z^n$, $n \geq 3$, and for SAW in $\Z^n$, $n \geq 2$, $\delta_N$ remains strictly less than one, which means that the loss of
information due to the coarse graining becomes substantial (see
(\ref{eq2.3b}) and Theorem~\ref{thmD}).

We also consider the problem addressed already in~\cite{Leb}, i.e.\
how the number of different \emph{physical} contact matrices $W(N)$
depends on the polymer length $N$. It was shown there analytically
that $W(N)$ increases exponentially in $N$, and numerically that the
growth exponent $\gamma_N$ is strictly less than the growth exponent for
the total number of SAW. In our work we give a rigorous proof of this
(see Theorem~\ref{thmD}). We  also consider the same problem for
SRW. Surprisingly, the growth exponent for the contact matrices is now
the same as for the 
number of SRW in all dimensions. The reason for this is that
the probability distribution of the number of distinct visited sites
(divided by $N$) has a left tail which does not decay exponentially fast
in $N$. To conclude, we provide lower bounds on $\gamma_N$, relevant
to the finite size behaviour (see (\ref{eq2.2a}) and (\ref{eq2.3a})).

The outline of the rest of the paper is as follows.
In section~\ref{sect1} we introduce the model, the relevant quantities and the studied examples.
The main results are presented and briefly discussed in section~\ref{sect2}.
Sections \ref{sect3}, \ref{sect4}, and \ref{sect5} are devoted to the proof of the main results.

\section{Preliminaries}\label{sect1}
We define more precisely the quantities and the examples which will be studied.
$\Omega_N$ is the set of all polymers containing $N+1$ monomers, where the configuration of such a polymer is specified by $\omega_N = (\omega(0),\omega(1),...,\omega(N))$ with $\omega(0)\equiv 0$ and $\omega(i+1)-\omega(i) = \pm e_\alpha$, where $e_\alpha$ is one of the unit directions on $\Z^n$, $\alpha = 1,...,n$. 
Let there be given some probability distribution $\Pb(\omega)$ on $\Omega$. (We shall drop the subscript $N$ whenever possible.)
The contact matrices $\Cset = \{\C(\omega)\}_{\omega\in \Omega}$ partition $\Omega$ into
sets $\Omega_C = \{ \omega: \C(\omega)=C\}$, with 
\begin{equation}\label{eq1.2}
\deg \C = |\Omega_C| \end{equation}
the number of configurations $\omega \in \Omega_C$.
The probability of $\omega$ being in $\Omega_C$ is then
\begin{equation}\label{eq1.3}
\Pb(C) = \sum_{\omega \in \Omega_C} \Pb(\omega).
\end{equation}
To measure the information lost in the coarse-graining we may compare the
Shannon entropy $S_C$ of the coarse grained measure $\Pb(C)$ with the
fine grained entropy $S$, 
\begin{equation}\label{eq1.4}
S = -\sum_{\omega \in \Omega} \Pb(\omega) \ln \Pb(\omega).
\end{equation}
We then have
\begin{eqnarray}\label{eq1.5}
S_C = -\sum_{C\in \Cset} \Pb(C) \ln \Pb(C)
& = &-\sum_{C\in \Cset} \sum_{\omega \in \Omega_C} \Pb(\omega) \ln
\bigg(\frac{\Pb(C)}{\Pb(\omega)} \Pb(\omega)\bigg)\nonumber \\
&=& S - \hat S_C
\end{eqnarray}
where
\begin{equation}\label{eq1.7}
\hat S_C = -\sum_{C \in \Cset} \Pb(C) \sum_{\omega \in \Omega_C} \Pb(\omega|C)
\ln \Pb(\omega|C)
\end{equation}
with 
\begin{equation}\label{eq1.8}
\Pb(\omega|C) = \Pb(\omega)/\Pb(C) \quad \textrm{for }\omega \in \Omega_C
\end{equation}
is the conditional probability of $\omega$ given that it is
in $\Omega_C$.  We can thus  think of $\hat S_C$ as the average
``conditional entropy'' relative to $\C$.  Since $\hat S_C \geq 0$ we
clearly have $S_C \leq S$, and 
$S - S_C$ is then a measure of information lost in the coarse
graining \cite{Bill}. The question is how much. In particular we may ask how does
\begin{equation}\label{eq1.9}
\delta_N = S_C/S 
\end{equation}
behave as $N \to \infty$.  

Before answering this question we note that 
\begin{equation}\label{eq1.9b}
S_C \leq \bar S_C = \ln |{\bf C}|,
\end{equation} where $\bar S_C$ is the entropy of the distribution
$\bar \Pb(C)$  which assigns equal weight to all $C\in\Cset$,
i.e.\ $\bar \Pb(C) = W(N)^{-1}$, with $W(N)\equiv|{\bf C}|$
the total number of different coarse grained components, i.e.\ 
contact matrices. We may then also define 
\begin{equation}
\gamma_N = \bar S_C/S
\end{equation}
and by (\ref{eq1.9b})
\begin{equation}
\delta_N \leq \gamma_N \leq 1.
\end{equation}
The last inequality is obtained by replacing $\Pb(C)$ by $\bar\Pb(C)$ in (\ref{eq1.5})-(\ref{eq1.8}) and using that the corresponding $\hat S_C$ is positive.

So far everything is completely general.  We shall now specialize to
the case where all permissible configurations $\omega$, i.e.\ 
all those for which $\Pb(\omega) \ne 0$, have the same probability.   
Then
\begin{equation}\label{eq1.10a}
W(N) = |\Cset| = \sum_{\omega\in \Omega} 1/ \deg \C(\omega)=|\Omega|
\E{(\deg \C)^{-1}},
\end{equation} 
where $\E$ is the expectation value with respect to the relevant uniform
distribution. 

The examples we shall consider here are:

1)  The weights are those of simple symmetric random
walks (SRW) on $\Z^n$, i.e.\ $|\Omega| = (2n)^N$ and $\Pb(\omega) = (2n)^{-N}$ for all $\omega$.

2) The polymers behave like self avoiding walks (SAW) on $\Z^n$, i.e.\ 
the configuration space 
$\Omega$ consists of all $\omega$ s.t.\ $\omega(i) \neq \omega(j)$ for $i
\neq j$, and $\Pb(\omega) = |\Omega|^{-1}$, where $|\Omega| \sim
\mu_{SAW}^N$ is the number of SAW on $\Z^n$ of length $N$.  SAW model the
steric exclusion effects of the monomers and are frequently used as a
model for polymers~\cite{Freed,LD}.

3) The chains behave like bond self avoiding walks (BAW) on $\Z^n$ in which
case $\Omega$ consists of all $\omega$ such that the pair $[\omega(i),
\omega(i+1)] \ne [\omega(j), \omega(j \pm 1)]$ for $i \neq j$, and
$\Pb(\omega) = |\Omega|^{-1}$, $|\Omega| \sim \mu_{BAW}^N$, the number of BAW \cite{Madras}.

Note that for uniform distributions $S$ is just the logarithm of the
total number of configurations, i.e.\ $S = \ln |\Omega|$, so
$\gamma_N$ is just the ratio of the logarithms of the numbers of
contact matrices and random walks.

The behavior of $\gamma_N$  was studied in~\cite{Leb} for the case of SAW, with
\begin{equation}\label{eqCSAW}
\C _{ij}(\omega)=\left\{
\begin{array}{l}
1 \quad \textrm{ if } \arrowvert\omega(i)-\omega(j)\arrowvert = 1, \, \arrowvert
i-j\arrowvert > 1,\\ 0 \quad \textrm{ otherwise.}
\end{array}\right.
\end{equation}
(The inequality can in fact be made an equality here since $1$ is the minimal distance between $\omega(i)$ and $\omega(j)$ for $i \neq j$.)
Numerical studies \cite{Leb} for $n=2$ indicated that $\gamma_N$
remains  strictly less than $1$ in the limit $N \to \infty$. It is then
natural to ask whether the same is true for the SRW when we again
define $\C_{ij}(\omega) = 1$ when $\omega(i)$ and $\omega(j)$ are as close as
they can be, i.e.\ when $a$ in (\ref{eq1.1}) is set equal to zero 
\begin{equation}\label{eq1.11}
\C_{ij}(\omega) = \left\{\begin{array}{ll}1 & \textrm{for }|\omega(i) -
\omega(j)| = 0,\, i \ne j \\ 0 & \textrm{otherwise}\end{array}\right.
\end{equation}
Note that for this case $W(N)$ satisfies
\begin{equation}\label{eq2.14}
W(N_1+N_2) \geq W(N_1) W(N_2).
\end{equation}
Since for SRW 
$\gamma_N = \frac{\ln W(N)}{N\ln (2n)}$ it follows from (\ref{eq2.14})
that $\gamma_N$ is monotone non-decreasing in $N$ and thus that
$\lim_{N\to\infty}\gamma_N$ exists:  remember $\gamma_N \leq 1$.

In the present work we prove some results about $\delta_N$ and
$\gamma_N$ for all the above examples. (Some of these generalize
readily to other uniform distributions.)

\section{Main Results}\label{sect2}
\begin{thm}\label{thmA}
For SRW on $\Z^2$, there exist constants $\kappa,\kappa_1,\kappa_2>0$ such that for
large $N$,
\begin{equation*}
\gamma_N \geq 1-\frac{\kappa \ln{N}}{N^{1/3}},\addtocounter{equation}{1}\tag{\theequation a}\label{eq2.2a}
\end{equation*}
\begin{equation*}
1-\frac{\kappa_1}{\ln{N}}\leq \delta_N\leq 1-\frac{\kappa_2}{(\ln{N})^2}.\tag{\theequation b}\label{eq2.2b}
\end{equation*}
Consequently, $\gamma_N$ and $\delta_N \to 1$, as $N \to \infty$.
\end{thm}

\begin{thm}\label{thmB}
For SRW on $\Z^n$, $n\geq 3$, there exist constants $\kappa_n, \kappa^\prime_n >0$ such that for $N$ large enough,
\begin{equation*}
\gamma_N \geq 1 - \frac{\kappa_n}{N^{2/{(n+2)}}}, \addtocounter{equation}{1}\tag{\theequation a}\label{eq2.3a}
\end{equation*}
\begin{equation*}
\delta_N \leq 1 - \kappa^\prime_n.\tag{\theequation b}\label{eq2.3b}
\end{equation*}
Hence, $\gamma_N\to 1$, as $N\to\infty$ while $\limsup_{N\to\infty}\delta_N < 1$.
\end{thm}

\begin{thm}\label{thmD}
For SAW and BAW, on $\Z^n$, $n \geq 2$, $\limsup_{N\to\infty}\gamma_N < 1$
and ipso-facto $\limsup_{N\to\infty}\delta_N < 1$.
\end{thm}

We are indebted to Harry Kesten for the key idea in the proof of
Theorem~\ref{thmD} for SAW. The extension to BAW is straightforward.

Intuitively we expect that the larger the number of intersections, the
more information is contained in the contact matrices.  
It is therefore not surprising that for recurrent RW, such as SRW in $n=2$, both
$\gamma_N$ and $\delta_N$ would go to $1$. (For the degenerate case
$n=1$ there are, for all the cases, just twice as many random walks as contact matrices corresponding to whether the first step is to the right or the left.)
One expects however that for RW which have a strong tendency to spread out such as SAW in $n\geq 2$, and SRW in $n \geq 3$ the contact matrices lose too much
information.  
This is indeed reflected in $\delta_N < 1$ for SAW in $n \geq 2$, and SRW in $n \geq 3$.  The same is true for $\gamma_N$ for SAW.  Surprisingly however $\gamma_N \to 1$ for SRW in all dimensions.  The reason for this, as we shall see, is that the probability that an SRW of length $N$ in $\Z^n$ visits $R_N \leq \epsilon N$ distinct sites goes to zero slower than exponentially when $N \to \infty$ for
any fixed $\epsilon > 0$.

The outline of the rest of the paper is as follows.
In section~\ref{sect3} we first present some general inequalities and then prove  the results about $\gamma_N$ for SRW.
In section~\ref{sect4} we give some bounds on the degeneracy of SRW
and then prove the results about $\delta_N$ for SRW.
In section~\ref{sect5} we prove Theorem~\ref{thmD} for SAW and BAW.

\section{Proof of Results for $\gamma_N$}\label{sect3}

\noindent {\bf Some Inequalities}

Using Jensen's inequality on the average over $\Omega_C$, gives 
\begin{eqnarray}
\sum_{\omega \in \Omega_C} \frac{1}{\deg \C} (\Pb(\omega) \deg \C) 
\ln(\Pb(\omega) \deg \C) &\geq &
\sum_{\omega \in \Omega_C} \Pb(\omega) \ln\bigg(\sum_{\omega \in \Omega_C}
\Pb(\omega)\bigg) \nonumber \\
&=& \Pb(C) \ln \Pb(C).
\end{eqnarray}
Writing now
\begin{equation}
S =-\sum_{C\in \Cset} \sum_{\omega \in \Omega_C} \Pb(\omega)
(\ln(\Pb(\omega)\deg \C) - \ln\deg \C)
\end{equation}
we obtain\footnote{The right side of (\ref{eq3.3}) is the
     maximum of the entropy over all measures, $\mu(\omega)$ such that
     $\mu(\C) = \Pb(\C)$.}
\begin{equation}\label{eq3.3}
S \leq S_C + \E(\ln \deg \C),
\end{equation}
which yields
\begin{equation}\label{eq3.5}
1 - \frac{\E(\ln\deg \C)}{S} \leq \delta_N \leq 1.
\end{equation}

We next give an upper bound for the degeneracy of the contact matrices defined in (\ref{eq1.11}). This is the key to our results for SRW. Note that for the examples considered here, the first inequality in (\ref{eq3.5}) is indeed an equality. 

Define the \emph{range} $R_N$ of a SRW with length $N$ to be \emph{the number of distinct sites visited by the walk}. 

\begin{lem}\label{lemma3.1}
Let $\omega$  have a 
range $R_N = M$ and let $\C(\omega)$ be its contact matrix. Then
\begin{equation}\label{eq3.6}
\deg\C (\omega) \leq (2n)^M.
\end{equation}
\begin{proof}[\sc{Proof}] The contact matrix
$\C(\omega)$ has $N+1-M$ columns with ``1'''s in the upper
triangular part, because we have $N+1-M$ intersections. Let us now
construct all random walks $\omega'$ such that
$\C(\omega')=\C(\omega)$. Consider now $\omega'(k)$
with $k \neq 0$, then there are two possible cases:\\[6pt]
1. there exists a $i<k$ such that $\C_{ik}(\omega)=1$,\\[6pt]
2. for all $i<k$, $\C_{ik}(\omega)=0$.

In the first case, $\omega'(k) = \omega'(i)$ and therefore we have only one choice for it. In the second case, the $k^{th}$ step will occupy a place which was never occupied before. Therefore we have \emph{at most} $2n$ possibles choices.

For a $\omega'$ with a contact matrix $\C(\omega')=\C(\omega)$ there are $M-1$ steps for which we are in the second case because the starting point is fixed
at the origin and $N+1-M$ steps for which we are in the first one.
Therefore there are \emph{at most} $(2n)^{M-1}\leq (2n)^M$ different
$\omega'$ satisfying $\C(\omega')=\C(\omega)$,
i.e.\ $\deg\C(\omega) \leq (2n)^M.$
\end{proof}
\end{lem}

\begin{proof}[\sc{\bf{Proof of (\ref{eq2.2a})}}]
We first note that  $\delta_N, \gamma_N \to 1$ for general 
recurrent RW, which includes SRW on $\Z^2$.  For such
walks $\E(R_N)/N\to 0$ as $N\to\infty$, therefore using
(\ref{eq3.5}) and Lemma~\ref{lemma3.1},
\begin{equation}
\delta_N\geq 1-\frac{\E(\ln\deg \C)}{N\ln{2n}}\geq 1-\frac{\E(R_N)}{N}\to
1\textrm{ as }N\to\infty.
\end{equation}

To prove (\ref{eq2.2a}) consider the subset $\Omega^\alpha$ of SRW on $\Z^2$
defined as
\begin{equation}
\Omega^\alpha=\{\omega \in \Omega_N \st \omega(k\cdot 4[N^\alpha])=0,
k=0,1,\ldots,k_{max}\}.
\end{equation}
where $k_{max}$ is the largest integer $k$ such that $k 4[N^\alpha] \leq
N$. Let us take $0< \alpha < 1/2$. Each $\omega\in \Omega^\alpha$ returns to the origin after $4[N^\alpha]$ steps, it is therefore contained in a cube of edge length $4[N^\alpha]$ (except eventually for the last $2[N^\alpha]$ steps).
Using Stirling formula we have, for some $K>0$,
\begin{equation}
\Pb(\omega(4M)=0)\geq \frac{(4M)!}{(M!)^4 4^{4M}}\geq K M^{-3/2}
\end{equation}
because $\{\omega(4M)=0\} \supset \{\omega(4M)=0\textrm{ with }M\textrm{ steps
in each direction}\}$. Then
\begin{eqnarray}\label{eq3.10}
\Pb(R_N\leq N^\beta\equiv 4^2 [N^\alpha]^2+2[N^\alpha])&\geq &\Pb(\omega \in
\Omega^\alpha)\nonumber \\ &\geq& (K/[N^\alpha]^{3/2})^{N/4[N^\alpha]}.
\end{eqnarray}
Therefore combining (\ref{eq1.10a}), (\ref{eq3.6}) and (\ref{eq3.10}) for $n = 2$ obtain 
\begin{equation}
W(N)\geq 4^N \Pb(R_N\leq N^\beta)/4^{(N^\beta)},
\end{equation} which implies
\begin{equation}\label{eq3.10b}
1\geq \gamma_N \geq
1-\kappa\frac{\ln{N}}{[N^\alpha]}-\frac{4^2[N^\alpha]^2+2[N^\alpha]}{N},
\end{equation}
for a $\kappa>0$. For $\alpha=1/3$ the RHS of (\ref{eq3.10b}) is optimized and
the term with the logarithm dominates the last one. This proves the bound for
$n=2$.
\end{proof}

\begin{proof}[\sc{\bf{Proof of (\ref{eq2.3a})}}]
For $n > 2$, $\Pb(R_N = X) \sim \exp (-aN/X^{2/n})$ when $X \to
\infty$,  $\frac{X}{N} \to 0$ (see \cite{Nieu} and pp.\ 88-92 of \cite{WijlandThese}). Therefore for
$\alpha \in (0,1)$, $\Pb(R_N = N^\alpha) \sim \exp(-aN^{1-2\alpha/n}).$
But for an $\omega$ with $R_N=M$,
$\deg\C(\omega)\leq(2n)^M$, see Lemma~\ref{lemma3.1}.
Therefore restricting the sum in (\ref{eq1.10a}) to $\omega
\in \Omega^\alpha_N$, we have $W(N) \geq (2n)^N \Pb(R_N
= N^\alpha)/(2n)^{N^\alpha}$, since the numerator is just the number of
terms in that sum.  This implies, for large $N$, 
\begin{equation}
1 \geq \gamma_N \geq 1 - \sigma(N,\alpha)
\end{equation}
where $\sigma(N,\alpha)
= N^{\alpha-1}+a N^{-2\alpha/n}/\ln{2n}.$
Choosing $\alpha \in (0,1)$ which minimizes $\sigma(N,\alpha)$
for large $N$, we obtain $\alpha-1=-2/(n+2)$.
Taking $\kappa_n=(\ln{2n}+a)/\ln{2n}$ completes the proof.
\end{proof}
\noindent This theorem implies that for $N$ large we have (up to smaller corrections),
\begin{equation}
W(N) \geq (2n)^N (2n)^{-\kappa_n N^{n/{(n+2)}}}.
\end{equation}

There exists also an upper bound on $\gamma_N$ which depends on the decrease of $\Pb(R_N/N<\varepsilon)$.
\begin{prop}\label{bornesupgamma}
For all fixed $\varepsilon>0$, there exists a constant $\kappa'>0$ such
that for $N$ large enough, 
\begin{equation}\label{prop3.6}
\gamma_N \leq 1 - \kappa'\frac{\arrowvert\ln{\Pb(R_N/N \leq\varepsilon)}\arrowvert}{N}.
\end{equation}
\end{prop}
\noindent The outline of the proof will be given in the Appendix.

\section{Bounds on the degeneracy of SRW}\label{sect4}
\subsection{SRW on $\Z^2$}
For SRW on $\Z^2$, $R_N \sim \frac{\pi N}{\ln{N}}$, more precisely (see e.g.\ 
\cite{Wijland2d}),
\begin{equation}\label{eq4.1}
\E(R_N)=\frac{\pi N}{\ln{8N}}\left(1+\mathcal{O}\left(1/\ln{N}\right)\right).
\end{equation}

Next we apply a result of van Wijland, Caser and Hilhorst \cite{Wijland2d}.
Let the \emph{support of} $\omega$ be defined to be \emph{the set of points visited by} $\omega$.
Consider two finite disjoint sets of lattice points
$A_u$ and $A_v$. The ``pattern'' centered at ${\bf x}$ associated with the sets
$A_u$ and $A_v$ is a configuration of $|A_v|$ visited sites ${\bf x+z}$,
${\bf z}\in A_v$ and of $|A_u|$ unvisited sites ${\bf x+z'}$, ${\bf z'} \in
A_u$. We say that the pattern appears in the support of $\omega$ at ${\bf x}$ if
the lattice points ${\bf x+z}$, ${\bf z}\in A_v$, are in the support of $\omega$
and the lattice points ${\bf x+z'}$, ${\bf z'} \in A_u$, are not in the support
of $\omega$. The numbers of times that a pattern appears in the support of $\omega$ is
then the number of different ${\bf x}\in \Z^n$ such that it appears at ${\bf x}$.

\begin{figure}[t]
\begin{center}
\psfrag{x}[0][0][1.2]{$x$}
\psfrag{y}[0][0][1.2]{$y$}
\includegraphics[angle=0,height=4cm]{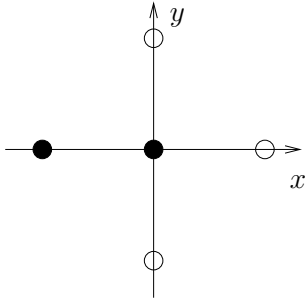}
\end{center}
\caption{The pattern $Q$. The visited sites of $Q$ are black.\label{figP2d}}
\end{figure}

Let us consider the ``pattern $Q$'' defined as the set composed of the following
sets $A_v$ and $A_u$:
$A_v(Q)=\{(0,0), (-1,0) \}$ and $A_u(Q)=\{(1,0), (0,-1), (0,1) ,
(-1,1)\}$ (see Figure~\ref{figP2d}).

Let $Q_N=Q_N(\omega)$ be the number of times that $Q$ appears in
the support of $\omega$. Then using \cite{Wijland2d},
$\E(Q_N)=\frac{\pi^2N}{(\ln{8N})^2}m_1+\mathcal{O}\left(\frac{N}{(\ln{8N})^3}\right)$ where $m_1=m_1(Q)$ is a constant, and $Q_N-\E(Q_N) \simeq
\frac{2\mathcal{A}}{\ln{8N}}\E(Q_N)\gamma(N)$ where $\gamma(N)$ is
a random variable (Varadhan's renormalized local time of self-intersections, see~\cite{Gal}) with mean $0$ and variance $1$. $\mathcal{A}$ is a constant
given in~\cite{Wijland2d} whose value is $\sim 1.3034$. We computed $m_1$ finding $m_1\cong 2.78\cdot10^{-3}.$

These results imply the following proposition.
\begin{prop}\label{prop4.3}
For simple random walks on $\Z^2$, there exists a $\nu > 0$ such that
\begin{equation}\label{eq4.2}
\lim_{N \to \infty}{\Pb\left(\deg\C (\omega)
\geq e^{\nu N/(\ln{N})^2}\right)}=1.
\end{equation}
\begin{proof}[\sc{Proof}]
Suppose that the pattern $Q$, centered at  $\zeta \in \Z^2$, exists in the support of a random walk $\omega$. Let us consider the following transformation:  
\begin{equation}
\begin{array}{rcl}
T_\zeta: \Omega_N & \longmapsto & \Omega_N \nonumber \\
\phantom{T_\zeta: }\omega & \longrightarrow & T_\zeta(\omega)  = \left\{
\begin{array}{ll}
\omega(i) & \textrm{ if } \omega(i) \neq \zeta, \\
\zeta+(-1,1) & \textrm{ if } \omega(i) = \zeta. \\
\end{array} \right. 
\end{array}
\end{equation}
In other words we exchange the points $\zeta$ and $\zeta+(-1,1)$. This
application does not change the contact matrix of the random walk,
because $\zeta+(-1,1)$ is connected only with $\zeta+(-1,0)$.
We have to prove that the probability of having  the
pattern $Q$ in the support of a random walk at least $M=\nu N /(\ln{N})^2$ times goes to $1$ as $N\to\infty$.
A RW with $M$ times the pattern $Q$ appearing in its support is at
least $2^M$ times degenerate: we can apply or not apply $T_\zeta$ independently
for each $\zeta$ such that $Q$ appears in the support of $\omega$ (centered in $\zeta$).

We want an upper bound of $\Pb\left(Q_N < \alpha \frac{\mu N}{(\ln{N})^2}\right)$ for $\alpha \in (0,1)$ and $\mu = m_1 \pi^2$. For each $k>0$ and $N$ large enough,
\begin{equation}
\Pb\bigg(Q_N <
\alpha \frac{\mu N}{(\ln{N})^2}\bigg) \leq \Pb\bigg(Q_N-\E(Q_N) \leq -k a_Q
\frac{N}{(\ln{8N})^3}\bigg)
\end{equation}
with $a_Q=2\mu \mathcal{A}$.
In fact, for $N$ large enough, $\E(Q_N)= \frac{\mu N}{(\ln{8N})^2}+ \mathcal{O}(N/(\ln{8N})^3)$
and therefore for each $\alpha < 1$, $\frac{\mu N}{(\ln{8N})^2} +\mathcal{O}(N/(\ln{8N})^3)-k
a_Q \frac{N}{(\ln{8N})^3} \geq \alpha \frac{\mu N}{(\ln{N})^2}$.
Thus
\begin{eqnarray}
\Pb\left(Q_N < \alpha\frac{\mu N}{(\ln{N})^2}\right) & \leq & \Pb\left(Q_N-\E(Q_N)
\leq -k a_Q \frac{N}{(\ln{8N})^3}\right) \nonumber \\
& \leq & \frac{\E\left(Q_N-\E(Q_N)\right)^2}{k^2 a_Q^2 \frac{N^2}{(\ln{8N})^6}}
\stackrel{N \to \infty}{\longrightarrow } \frac{1}{k^2}.
\end{eqnarray}
Therefore for each $\alpha \in (0,1)$ we have
$\forall \, k > 0$, 
\begin{equation}
\lim_{N \to \infty}{\Pb\left(Q_N<\alpha \frac{\mu
N}{(\ln{N})^2}\right) \leq \frac{1}{k^2}}.
\end{equation}
This implies that for all $\alpha \in (0,1)$, $\lim_{N \to
\infty}{\Pb\left(Q_N \geq \alpha \frac{\mu N}{(\ln{N})^2}\right)}=1.$
All the random walks with a number of pattern $Q$ in their support more than
$\alpha \frac{\mu N}{(\ln{N})^2}$ are more degenerate than $e^{\alpha \mu
N\ln{2}/(\ln{N})^2}$. Then, since $\alpha \in (0,1)$, for all
choice of $\nu < \mu \ln{2}$ we have (\ref{eq4.2}).
\end{proof}
\end{prop}

\subsection{SRW on $\Z^n$, $n \geq 3$}
Let us define, for $n=3$, the pattern $P$ as consisting of
a set $A_v$ of visited sites and a set $A_u$ of unvisited sites as follows:
$A_v(P)=\{(0,0,0), (-1,0,0)\}$ and
\mbox{$A_u(P)=\{(1,0,0), (0,1,0), (0,-1,0), (0,0,1),  (0,0,-1),
(-1,1,0)\}$}.

Let $P_N=P_N(\omega)$ be the number  of times that $P$ appears in the support of
$\omega$. Then using \cite{Wijland3d} we have $\E(P_N)=m_1 N +
\mathcal{O}(\sqrt{N})$ and $P_N-\E(P_N) \simeq a_P \sqrt{N\ln{N}}\eta(N)$
where $m_1=2.5\cdot10^{-3}$, $a_P=1.2\cdot10^{-2}$ and $\eta(N)$ is a random
variable with normal distribution $\mathcal{N}(0,1)$.
\begin{prop}\label{prop4.4}For simple random walks on $\Z^3$, there exists a $\nu > 0$ such
that
\begin{equation}\label{eq4.7}
\lim_{N \to \infty}{\Pb\left(\deg\C (\omega)
\geq e^{\nu N}\right)}=1.
\end{equation}
\begin{proof}[\sc{Proof}]
The proof is very close to the one of
Proposition~\ref{prop4.3}. This time we exchange the sites
$\zeta$ and $\zeta+(-1,1,0)$ (if $P$ appears centered in $\zeta$)
and we prove that $\forall k >0$ and $N$ large enough $\Pb(P_N <
\alpha m_1 N) \leq \frac{1}{k^2}$ if $\alpha<1$. Then for all choice of $\nu < m_1 \ln{2}$
(\ref{eq4.7}) holds.
\end{proof}
\end{prop}
In dimension $n\geq 4$ the same result (with a different value of $m_1$) is
expected to hold. In fact a similar pattern in $n\geq 4$ has $\E(P_N)=m_1
N + \mathcal{O}(\ln{N})$ in $n=4$, $\E(P_N)=m_1 N + \mathcal{O}(1)$ in $n\geq 5$
and $\E(P_N-\E(P_N))^2=K_P N+o(N)$, see~\cite{Wijland3d}.
The only point that one should prove for $n\geq 4$ is that $m_1 \neq 0$.

\subsection{Proof of (\ref{eq2.2b}) and (\ref{eq2.3b})}
The previous results on the degeneracy lead to the following results.

\begin{proof}[\sc{\bf{Proof of (\ref{eq2.2b})}}]
Using Propositions~\ref{prop4.3} and (\ref{eq4.1}) we obtain bounds on
$\delta_N$, for $n=2$.
\begin{equation}
\delta_N\leq 1-\Pb\left(\deg\C \geq e^{\nu N/(\ln{N})^2}\right)
\frac{\nu}{\ln{4}(\ln{N})^2}.
\end{equation}
By Proposition~\ref{prop4.3}, there exists a $\nu>0$ such that
$\Pb\left(\deg\C \geq e^{\nu N/(\ln{N})^2}\right)\to 1$ as
$N\to\infty$. This gives 
\begin{equation}
\lim_{N\to\infty}(1-\delta_N)(\ln{N})^2\geq \nu/\ln{4}.
\end{equation}
Consequently for $\kappa_2<\nu/\ln{4}$ and $N$ large enough,
$\delta_N\leq 1-\frac{\kappa_2}{(\ln{N})^2}$.

On the other hand,
\begin{equation}\label{eq4.8}
\delta_N\geq 1-\frac{\E(\ln\deg \C)}{N\ln{2n}}\geq 1-\frac{\E(R_N)}{N}.
\end{equation}
Then
\begin{equation}
\lim_{N\to\infty}(1-\delta_N)\ln{N} \leq \lim_{N\to\infty}\E(R_N)\frac{\ln{N}}{N}=\pi,
\end{equation}
consequently for $\kappa_1>\pi$ and $N$ large enough,
$\delta_N\geq 1-\kappa_1/\ln{N}$.
\end{proof}

\begin{proof}[\sc{\bf{Proof of (\ref{eq2.3b})}}]
For all $\nu>0$,
\begin{equation}\label{eq6.9}
\delta_N \leq 1- \frac{\Pb(\deg \C  \geq e^{\nu N})}{\ln{(2n)}}\nu.
\end{equation}
By Proposition~\ref{prop4.4}, there exists, for $n \geq 3$, a $\nu>0$ such that 
$\Pb(\deg \C  \geq e^{\nu N}) \to 1$ as $N\to\infty$. Therefore for $N$ large enough 
\begin{equation}
\delta_N \leq 1-\frac{\nu/2}{\ln{(2n)}}<1,
\end{equation}
and $\limsup_{N\to\infty}\delta_N <1.$
\end{proof}

\section{Proof of Theorem~\ref{thmD}}\label{sect5}
The contact matrix of $\omega$ is now defined by (\ref{eqCSAW}).

Let us consider the case of SAW. We introduce some notation:
we consider a cube $D = \{ x \in \Z^n \st c_i \leq x^{(i)} \leq c_i+b, 1 \leq
i \leq n\}$ for some \mbox{$c=(c_1,\ldots,c_n) \in \Z^n$} with its boundary
$\partial{D} = \{ x \in \Z^n \st x^{(i)} = c_i+b \textrm{ or } x^{(i)} =
c_i, 1 \leq i \leq n\}$.
A path $P$ is a SAW of finite length, say $k$, starting at the origin, i.e.\ 
$P =\{X_i(P), 0 \leq i \leq k\}$ with $X_0(P)=0$.

We consider only paths such that there exists a cube $D$ with $X_0(P)=0$ and $X_k(P)$ two of its vertices and $X_j(P)\in D$ for all $0\leq j\leq k$. We say that $(P,D)$ occurs at the $r^{th}$ step $\omega$ if\\[6pt]
1. $X_{r+j}(\omega)-X_r(\omega)=X_j(P)$ for all $j=0,\ldots,k$ and\\[6pt]
2. $\omega$ does not occupy any other points of $D$.\\[6pt]
$\chi_N(j,(P,D))$ is the number of $\omega\in\Omega_N$ such that $(P,D)$ occurs \emph{at
most} at $j$ steps.

\begin{thm}[Kesten Pattern Theorem~\cite{Kesten}]\label{thm5.1}
Let $P$ be a SAW and $D$ a cube such that $D$ has $0$ and $X_k(P)$ as two of its vertices and contains $P$. Then
\begin{equation}
\limsup_{N \to \infty}{\left(\frac{\chi_N(a N,(P,D))}{\Card{\Omega_N}}\right)^{1/N}}<1\textrm{ for some }a>0
\end{equation}
where $\Card{\Omega_N}$ is the total number of SAW of length N.
\end{thm}
It is known that $\Card{\Omega_N}\simeq \mu_{SAW}^N$ with $\mu_{SAW}>0$.

\begin{proof}[\sc{\bf{Proof of Theorem~\ref{thmD}}}]
Let us take $b>2$ and consider a SAW path of length $k+2n$ constructed as
follows. The firsts $n$ steps of $P$ connect the points $(0,\ldots,0)$ and
$(1,\ldots,1)$. The following $k$ steps connect the points $(1,\ldots,1)$ and
$(b-1,\ldots,b-1)$ with a SAW remaining always in $D \setminus \partial{D}$.
The last $n$ steps of $P$ connect the points $(b-1,\ldots,b-1)$ and $(b,\ldots,b)$.
Let us divide the set $\Omega_N$ into a sum of two disjoint parts:
$\Omega_N=\Omega_N^a \cup (\Omega_N^a)^c$ where $\Omega_N^a =
\{ \omega \in \Omega_N \st (P,D) \textrm{ occurs at most } a
N\textrm{ times}\}$ and $(\Omega_N^a)^c$ its complementary set.
It follows from Theorem~\ref{thm5.1}  that \[\exists \, \zeta > 0
\st \Pb(\omega \in \Omega_N^a) \leq e^{- \zeta N}.\] Let us
take a $\omega \in (\Omega_N^a)^c$. Then $(P,D)$ occurs at
least $a N$ times in $\omega$. Consider an occurrence of $(P,D)$ in the piece
of $P$ between its $t^{th}$ and its $(t+k+2n)^{th}$ steps. We
apply an axis rotation of $2 \pi/n$ degrees to the cube
$D\setminus \partial{D}$, where the axis is its diagonal of
direction $(1,\ldots,1)$.  This transformation does not change the
contact matrix, and we can apply it $n$ times obtaining each time
a different SAW. For the chosen $\omega$ it can be done independently
in at least $a N$ different places, therefore the corresponding contact
matrix is at least $n^{a N}$ times degenerate.

Now we have an upper bound for the total number of contact matrices:
\begin{eqnarray}
W(N) &\leq &\Pb(\omega \in \Omega_N^a)\Card{\Omega_N}+\Pb(\omega \in
(\Omega_N^a)^c)\Card{\Omega_N} n^{-a N} \\
& \leq & \left( e^{- \zeta N}+ e^{- N a \ln{n}}\right) \Card{\Omega_N}. \nonumber
\end{eqnarray}
Defining $\alpha_M = \max\{\zeta,a\ln{n}\} >0$ and
$\alpha_m = \min\{\zeta,a\ln{n}\} >0$, we obtain
\begin{eqnarray}
\limsup_{N \to \infty}\gamma_N&\leq &
\lim_{N \to \infty}{\frac{\ln\left( e^{- \alpha_m N}
\left(1+ e^{- \frac{\alpha_M}{\alpha_m} N}\right) \right)+
\ln{\Card{\Omega_N}}}{\ln{\Card{\Omega_N}}}} \\
& = & 1-\frac{\alpha_m}{\ln{\mu_{SAW}}} < 1. \nonumber
\end{eqnarray}
\end{proof}

Now we consider the case of bond-self-avoiding walks (BAW).
We introduce some notation:
we consider a cube $D$ as for SAW and a cube $D^1= \{ x \in \Z^n \st c_i-1 \leq x^{(i)} \leq
c_i+b+1, 1 \leq i \leq n\}$ for some \mbox{$c=(c_1,\ldots,c_n) \in \Z^n$}.
In this case a path $P$ is a BAW instead of a SAW with the same conditions as
for SAW.
We consider only the paths  such that there exists a cube $D$ with $X_0(P)=0$ and $X_k(P)$ two of its vertices and
$X_j(P)\in D^1$ for all $0\leq j\leq k$.
We say that $(P,D)$ occurs at the $r^{th}$ step $\omega$ if\\[6pt]
1. $X_{r+j}(\omega)-X_r(\omega)=X_j(P)$ for all $j=0,\ldots,k$ and\\[6pt]
2. $\omega$ does not occupy any other points of $D$.\\[6pt]
$\chi_N(j,(P,D))$ is the number of $\omega\in\Omega_N$ such that $(P,D)$ occurs \emph{at most} at $j$ steps.
Theorem~\ref{thm5.1} holds also for BAW~\cite{Ferrari}.

\begin{prop}\label{prop5.3}
For BAW
\begin{equation}
\limsup_{N \to\infty}\gamma_N<1.
\end{equation}
\begin{proof}[\sc{Proof}]
The proof is identical to the one of Theorem~\ref{thmD}.
\end{proof}
\end{prop}

\begin{rem}
As noticed by Kesten in \cite{Kesten},
Theorem~\ref{thm5.1} could be proven also for other lattices in almost the
same way, therefore Theorem~\ref{thmD} should hold for other lattices
than $\Z^n$.
\end{rem}

\section*{Acknowledgments}
We thank H. Kesten for supplying us with the argument needed to prove
Theorem~\ref{thmD} for SAW, and S. Goldstein for useful discussions.
Research supported by NSF Grant DMR 98-13268, AFOSR Grant AF
49620-01-1-0154, and DIMACS and its supporting agencies, the NSF under
Contract No. STC-91-19999 and the N.J. Commission on Science and
Technology. Work of P.L. Ferrari partially supported by the Swiss
fellowship Sunburst-Fonds.

\appendix
\section{Outline of the proof of Proposition~\ref{bornesupgamma}}\label{appendix A}
Let $I_N=N+1-R_N$ be the number of intersections.
Consider an interval $J=[k_0,k_1]\subset [0,1]$ and the subset  
$\Lambda_N(J) = \{\omega \in \Omega_N \st I_N(\omega)/N \in J\}.$
We define the mean degeneracy on $\Lambda_N(J)$ by
$\langle\deg\C\rangle_J =\Card{\Lambda_N(J)}/W(N)_J,$
where $W(N)_J$ is the number of contact matrices corresponding to RW with $I_N/N\in J$.We set $d(J) = \liminf_{N \to\infty}{\frac{1}{N}\ln{\langle\deg\C\rangle_{J}}}.$
\begin{thm}\label{thm4.7}
For SRW on $\Z^n$, $n\geq 2$, and $\pi'=\lim_{N\to\infty}{\E(I_N)/N}$,
\begin{equation}
d(J=[k_0,k_1]) > 0 \textrm{ for all }k_0 < \pi'\textrm{ and } k_0< k_1 < 1.
\end{equation}
\end{thm}
For an $\omega \in \Omega_N$, let us define $F(\omega)$ to be the number of loops of length 4 which do not intersect the remaining part of $\omega$ (called
``free-4-loops'').
\begin{prop}\label{expdeg}
Let $J$ be as in Theorem~\ref{thm4.7}. Then there exists an $\alpha_J>0$ such that
\begin{equation}
\beta_J=\liminf_{N\to\infty}{-\frac{1}{N}\ln{\Pb\{\omega\in\Lambda_N(J)
\st F(\omega)\leq \alpha_J N\}}}>0.
\end{equation}
\end{prop}

\begin{proof}[\sc{\bf{Proof  of Theorem~\ref{thm4.7}}}]
For $k_0 < \pi'$ and $k_1 \in (k_0,1)$,
\begin{eqnarray}& &\hspace{-24pt}\frac{W(N)_J}{\Card{\Lambda_N(J)}}
\leq \Pb\{ F(\omega) \leq \alpha_J N \textrm{ for }\omega\in\Lambda_N(J) \} \\
&+&2^{-\alpha_J N}\Pb\{ F(\omega) > \alpha_J
N \textrm{ for }\omega\in\Lambda_N(J)\} \leq 2 \exp(-\min\{\beta_J,\alpha_J\ln{2}\}N)\nonumber
\end{eqnarray}
since a contact matrix with $M$ free-4-loops is at least $2^M$ times
degenerate. Then it follows by Proposition~\ref{expdeg} that 
$d(J)\geq \min\{\beta_J,\alpha_J \ln{2}\} > 0$.
\end{proof}

\noindent{\bf Outline of the proof of Proposition~\ref{expdeg}:}
Divide $\Z^n$ into disjoint $n$-cubes of edgelength 4. First we remark that at
least $aN$ cubes are visited by $\omega\in \Lambda_N(J)$, $a=(1-k_1)/4^n$,
and at least $aN/2$ are visited at most by $2/a$ steps.
Consider $\Lambda_N^{\alpha N}(J) =
\{\omega \in \Lambda_N(J) \st F(\omega)\leq \alpha N\}$,
$\alpha\ll 1$. We do two successive operations on $\omega\in \Lambda_N^{\alpha
N}(J)$.

1) We modify the free-4-loops so that the new RW $\widetilde \omega$ has
$F(\widetilde\omega)=0$. This is obtained by moving the 3rd step to the position of the 1st step of the free-4-loops.

2) We choose $2\alpha N$ cubes out of the first $aN/2$ visited less
than $2/a$ steps. The choice can be made in $\binom{a N/2}{2\alpha N}$ different ways. $\widetilde \omega$ passes in a cube not more that $2/a$ times and at each time we replace the path inside the chosen cubes by another one of length increased by 2 which remains on the boundary of the cube and leaving the enter and exit points unchanged. Therefore the center of the cubes are now empty. Secondly we add a free-4-loop in the center of the cubes the first time that are visited by $\widetilde \omega$.

The final RW have length $n \in [N(1+20\alpha), N(1 + c_2\alpha)]$ and
$I_n/n \leq k_1+c_2\alpha$, $c_2=17+8/a$. Then using some results of
Hamana and Kesten on $\psi(k)=\lim_{N\to\infty}-\frac{1}{N}\ln{\Pb(R_N/N\geq
k)}$ \cite{Hamana}, we conclude that, if $\beta_J=0$, for $\alpha$ small enough the number of
constructed RW exceeds the total number of RW with 
$n \in [N(1+20\alpha), N(1 + c_2\alpha)]$ and $I_n/n \leq k_1+c_2\alpha$.
Therefore $\beta_J>0$.

\noindent{\bf Outline of the proof of proposition~\ref{bornesupgamma}:}
$\Pb(R_N/N\leq \e)$ is not exponentially small in $N$ (see e.g.\ \cite{Hamana}
and proof of (\ref{eq2.2a})). Let $J_1=[0,1-\e)$ and
$J_2=[1-\e,1]$. Since $d(J_1)>0$, $W(N)_{J_1}$ is exponentially small compared with
$W(N)_{J_2}$ for $N$ large enough. Therefore for large $N$, 
$W(N)\simeq W(N)_{J_2}\leq \Card{\Omega_N} \Pb(R_N/N \leq \e)$, from which
follows (\ref{prop3.6}).

The complete proof can be found at\\ {\it
http://www-m5.ma.tum.de/pers/ferrari/homepage/download/appendix.ps.gz}.

\end{document}